\documentclass[preprint2]{aastex}

\slugcomment{}

\shorttitle{Ab Initio Equation of State for Giant Planets}
\shortauthors{Militzer and Hubbard}

\begin{document}

%\title{Ab Initio Equation of State Calculations
% of Hydrogen-Helium Mixtures and Application to the Mass-Radius Relationship of Solar and % Extrasolar Giant Planets}

\title{Ab Initio Equation of State 
for Hydrogen-Helium Mixtures with Recalibration of the Giant-Planet Mass-Radius Relation }

\author{B. Militzer}
\affil{Department of Earth and Planetary Science, Department of Astronomy, University of California,
    Berkeley, CA 94720, USA.}

\and

\author{W. B. Hubbard}
\affil{Lunar and Planetary Laboratory, The University of Arizona, Tucson, AZ 85721, USA.}
%\email{aastex-help@aas.org}

\begin{abstract}
  Using density functional molecular dynamics simulations, we
  determine the equation of state for hydrogen-helium mixtures
  spanning density-temperature conditions typical of giant planet
  interiors, $\sim$ 0.2$-$9$\,$g$\,$cm$^{-3}$ and 1000$-$80\,000$\,$K
  for a typical helium mass fraction of 0.245. In addition to
  computing internal energy and pressure, we determine the entropy
  using an {\it ab initio} thermodynamic integration technique. A
  comprehensive equation of state (EOS) table with 391
  density-temperature points is constructed and the results are
  presented in form of two-dimensional free energy fit for
  interpolation. Deviations between our {\it ab initio} EOS and the
  semi-analytical EOS model by Saumon and Chabrier are analyzed in
  detail, and we use the results for initial revision of the inferred
  thermal state of giant planets with known values for mass and
  radius. Changes are most pronounced for planets in the Jupiter mass
  range and below. We present a revision to the mass-radius
  relationship which makes the hottest exoplanets
  increase in radius by $\sim$ 0.2
  Jupiter radii at fixed entropy and for masses greater than $\sim$ 0.5
  Jupiter mass.  This change is large enough to have possible implications
  for some discrepant ``inflated giant exoplanets''.
\end{abstract}


\section{Introduction}

The semi-analytical model by~\citet{SC92} (SC) for the EOS of hydrogen
and its extension to hydrogen-helium mixtures~\citep{SC95} were very
successful and have been used in numerous calculations for the
interiors of giant planets. However, with the development of {\it ab
  initio} computer simulation techniques many uncontrolled
approximations can now be avoided, simplifications inherent to
analytical EOS models and severely limiting their predictive
capabilities in the regime of high density and low temperature where
interactions between particles are strong. Relying solely on
analytical methods, it is difficult to determine the ionization state
of the different chemical species that are present in the dense fluid.

{\it Ab initio} simulations allow one to study a fully interacting
system of particles and to determine its properties by deriving the
electronic states explicitly for every configuration of nuclei. No
parameters are adjusted to match experimental data, but {\it ab initio}
simulations still rely on approximations to solve the Schr{\"o}dinger
equation. However, they are not specific to the particular material
nor the pressure-temperature conditions under consideration.

In this paper, we rely on density functional molecular dynamics
(DFT-MD) simulations that have been employed before to study
hydrogen~\citep{Le00,Mi01,Desjarlais2003,Bonev2004,NHKFRB,Morales2010,Caillabet2011,Collins2012,Nettelmann_2012},
helium~\citep{Mi06,StixrudeJeanloz08,Mi09} and hydrogen-helium
mixtures~\citep{Vo07,Vo07b,MHVTB,Mi09,Hamel2011}. 
While the computation of the pressure and the internal energy is
straightforward from DFT-MD simulations, the entropy is not directly
accessible. However, an accurate knowledge of the entropy of
hydrogen-helium mixtures at high pressure is of crucial importance for
the determination of the temperature profile, the density, and the
thermal energy budget in the interior of a giant planet. In 2008, two
groups constructed Jupiter interior models from DFT-MD
simulations~\citep{MHVTB,NHKFRB}. While the derived pressures and
internal energies can be considered to be more reliable than those
predicted by the SC model, both papers predicted very different
interior temperature profiles for Jupiter~\citep{MH08b}. Using {\it ab
  initio} thermodynamic integration techniques (TDI), we recently
showed~\citep{Militzer2013}, that the work by~\citet{NHKFRB}
overestimated the temperature at Jupiter's core-mantle boundary (CMB)
by 3050 K (19\%) while we underestimated it by 2870 K
(18\%) in \citet{MHVTB}. The revised temperature for the Jupiter's CMB is
16$\,$150$\,$K and the corrections to the SC EOS model are in fact
only $-350\,$K.

At conditions of Jupiter's CMB, hydrogen is metallic and characterized
by a high degree of electronic degeneracy. Such a degenerate state is
described rather well by the the SC model. However, when we applied
the TDI technique to explicitly determine the entropy over a wide
range of pressure-temperature conditions, we identified a number of
discrepancies between the DFT-MD results and the SC model. Near the
molecular-to-metallic transition, our simulations predict a
significant shift of the adiabat towards higher densities. At high
temperature, where electronic excitations matter, our computed
entropies are higher than those of the SC model. We also do not
perfectly reproduce the SC entropies in the molecular regime at low
density.

Rather than providing a separate hydrogen and helium EOS and relying
on the linear mixing approximation~\citep{SC95,NHKFRB}, we computed
the EOS over a wide range of density-temperature conditions for a
representative mixing ratio of $N_{\rm He}=$18 helium atoms in $N_{\rm
  H}=220$ hydrogen atoms, corresponding to a helium mass fraction of
$Y$=0.245, which is close to the solar value. This means that the
nonideal mixing effects are fully incorporated. In~\citet{Vo07}, we
showed for example that the presence of helium makes the hydrogen
molecules more stable and reduces the dissociation fraction at given
pressure and temperature.  Even if other mixing ratios become of
interest, as the result of helium
rain~\citep{Stevenson77a,Morales2009,WilsonMilitzer2010,McMahon2012},
one is still better off by starting from an EOS for a typical
hydrogen-helium mixture and then perturbing the mixing ratio by a
comparatively small amount. Increasing or decreasing the helium
fraction requires knowledge of a helium or hydrogen EOS, respectively.
For the helium EOS, we recommend our first-principles
computation~\citep{Mi09} because it provides simulation data points
for the pressure, $P$, internal energy, $E$, Helmholtz free energy,
$F$, and entropy, $S$, over a wide parameter range and a
thermodynamically consistent free energy fit for interpolation. For
available hydrogen EOS work, we refer to the recent review by
~\citet{McMahon2012} but there has also been a considerable
theoretical effort compute the hydrogen EOS with semi-analytical
techniques~\citep{DW02,Kraeft2002,Rogers2002,Safa2008,Ebeling2012,Alastuey2012}.
If the perturbation in the helium fraction is sufficiently small, one
may use the SC EOS for simplicity.

\section{Ab Initio Simulations}

\begin{figure}
\epsscale{1.0}
\plotone{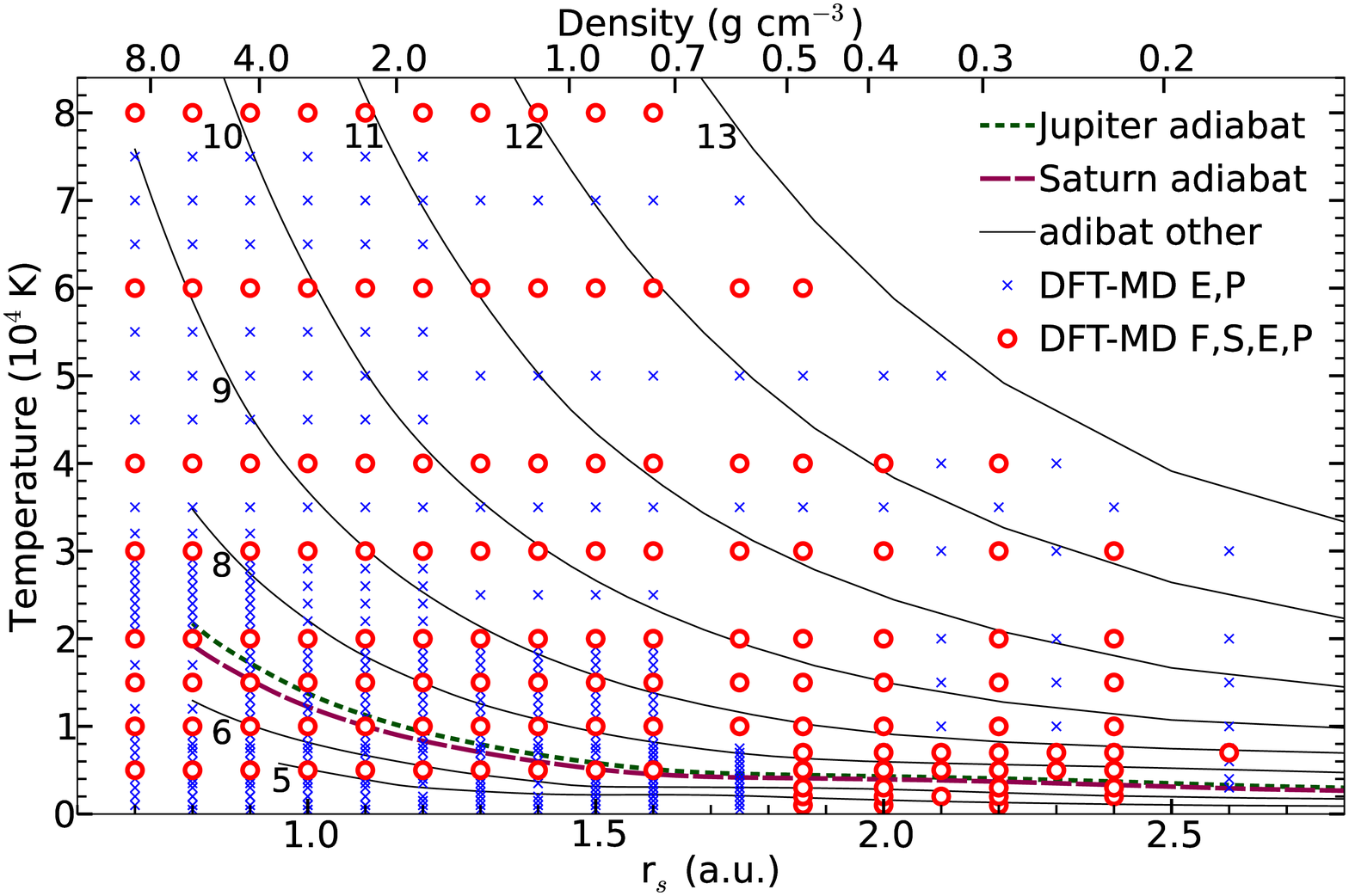}
\caption{Temperature-density conditions of DFT-MD
  simulations. The circles indicate parameters where entropy and free
  energy have been calculated in addition to the pressure and internal
  energy. The lines show adiabats. The labels specify their entropy
  values in units of $k_b$ per electron.\label{EOSTrs}}
\end{figure}

We base our {\it ab initio} entropy calculations on our recent
article~\citep{Militzer2013} where we showed how the TDI technique can
be extended to study molecular hydrogen and how it can be applied
efficiently to determine the entropy at high temperature where
electronic excitations matter. The TDI technique allows one to
determine the difference in the Helmholtz free energy between two
interacting many-body systems at fixed density and
temperature~\citep{Morales2009,WilsonMilitzer2010,WilsonMilitzer2012,WilsonMilitzer2012b,McMahon2012}.
We apply this method to determine the free energy difference between
the DFT simulations and a system of classical forces that we
construct:
\begin{equation}
F_{\rm DFT} - F_{\rm cl} = \int_0^1 d\lambda \; \left< V_{\rm KS} - V_{\rm cl} \right>_\lambda.
\label{tdint}
\end{equation}
The angle brackets represent an average over trajectories governed by
forces that are derived from a hybrid potential energy function,
$V_\lambda=\lambda V_{\rm KS}+(1-\lambda)V_{\rm cl}$. $V_{\rm cl}$ is
the potential energy of the classical system and $V_{\rm KS}$ is the
Kohn-Sham energy~\citep{KS65}. The presence of electronic excitations
leads to an intrinsic contribution to the entropy and affects the
forces on the nuclei~\citep{Wijs1998} that need to be derived from the
Mermin free energy~\citep{mermin}, $\Omega = V_{\rm KS} - TS_{\rm
  el}$. We combined both contributions into the following expression
for the {\it ab initio} entropy~\citep{Militzer2013}:
\begin{equation}
TS = \left< V_{\rm KS} \right> + \left< K_{\rm ion} \right>
- \int_0^1 d\lambda \; \left< \Omega - V_{\rm cl} \right>_\lambda - F_{\rm cl}.
\label{tdint2}
\end{equation}
$\left< V_{\rm KS} \right>$ includes contributions from partially
occupied excited states. The $\lambda$ integration was performed using
five independent MD simulations with $\lambda$ equally spaced between
0 and 1. To make this integration process efficient, we construct
the pair potentials of the classical system to match the DFT forces as
closely as possible~\citep{forcematching}. The computation of classical
free energy is performed with Monte Carlo methods by thermodynamic
integration to an system of noninteracting particles.

All simulations were performed with the VASP code~\citep{vasp1} with
pseudopotentials of the projector-augmented wave type~\citep{PAW} and
a plane wave basis set cutoff of at least 1000~eV. The
Perdew-Burke-Ernzerhof exchange-correlation functional~\citep{PBE} was
used throughout, but it has been shown recently that simulations based
on the local density approximation yielded very similar results for
Jupiter's deep interior~\citep{Militzer2013}. In the same article, we
also performed a combined finite-size and $k$ point analysis that
demonstrated that simulations with 256 electrons and the zone-average
point $k=(\frac{1}{4},\frac{1}{4},\frac{1}{4})$ are sufficiently
accurate. All results that we report in this article were thus
obtained with 220 hydrogen and 18 helium atoms in periodic boundary
conditions.

We used a MD time step 0.2 fs, except for temperature of
50$\,$000$\,$K and above where we used a time step of 0.1 fs to
accurately capture the more rapid collisions between particles at
elevated temperatures. All standard DFT-MD simulations that we
performed to determine $P$ and $E$ were 2.0 ps long, except at the
highest temperatures, where 1.0 ps were found to be sufficient because
the auto-correlation times are short and the error bars are small. All
simulations were initialized with positions and velocity vectors from
converged MD simulations at nearby densities and temperatures. This
allowed us to run the TDI simulations for only 0.5 ps at each
$\lambda$ point.

We also adjusted the number of orbitals in the calculations to
accommodate the partial occupation of excited electronic states
according to Mermin functional~\citep{mermin}. The number of orbitals
was increased until the error in the integral of the Fermi function
was reduced to less than 10$^{-5}$. This required many orbitals at
high temperature and low density. Up to 816 were used, a
significant increase in the computational cost over the 128 
needed for ground state calculations. This is the primary reason why we
omitted simulations that would lead to entropy values above approximately
12.5 $k_b$/el. The regime of higher temperatures can be studied much
more efficiently with path integral Monte Carlo (PIMC) simulations
because the computational cost of this alternative first-principles
simulation technique scales like $1/T$. PIMC simulations have been
applied to hydrogen~\citep{PC94,Ma96,Mi99,MC00,MG06,Hu2010,Hu2011},
helium~\citep{Mi06,Mi09}, and hydrogen-helium mixtures~\citep{Mi05} at
high pressure and temperature and most recently also to study the EOS
of carbon and water~\citep{DriverMilitzer2012}.

\section{Equation of State Results}

\begin{figure}
\epsscale{1.0}
\plotone{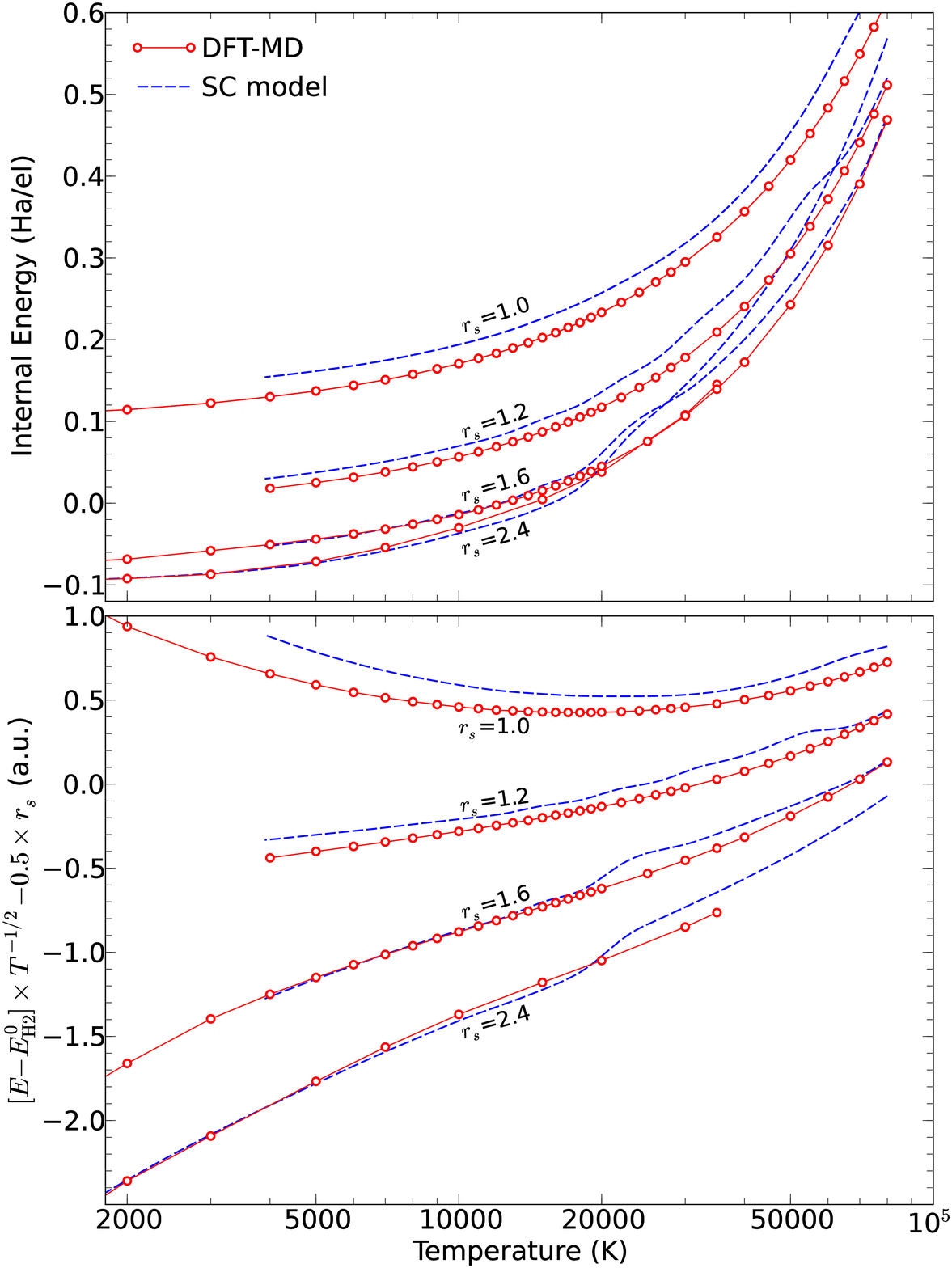}
\caption{Internal energy per electron as function of temperature for
  four different densities given in terms of $r_s$. Results from
  DFT-MD simulations are compared with the analytical SC model.
  \label{EOSET}}
\end{figure}

\begin{figure}
\epsscale{1.0}
\plotone{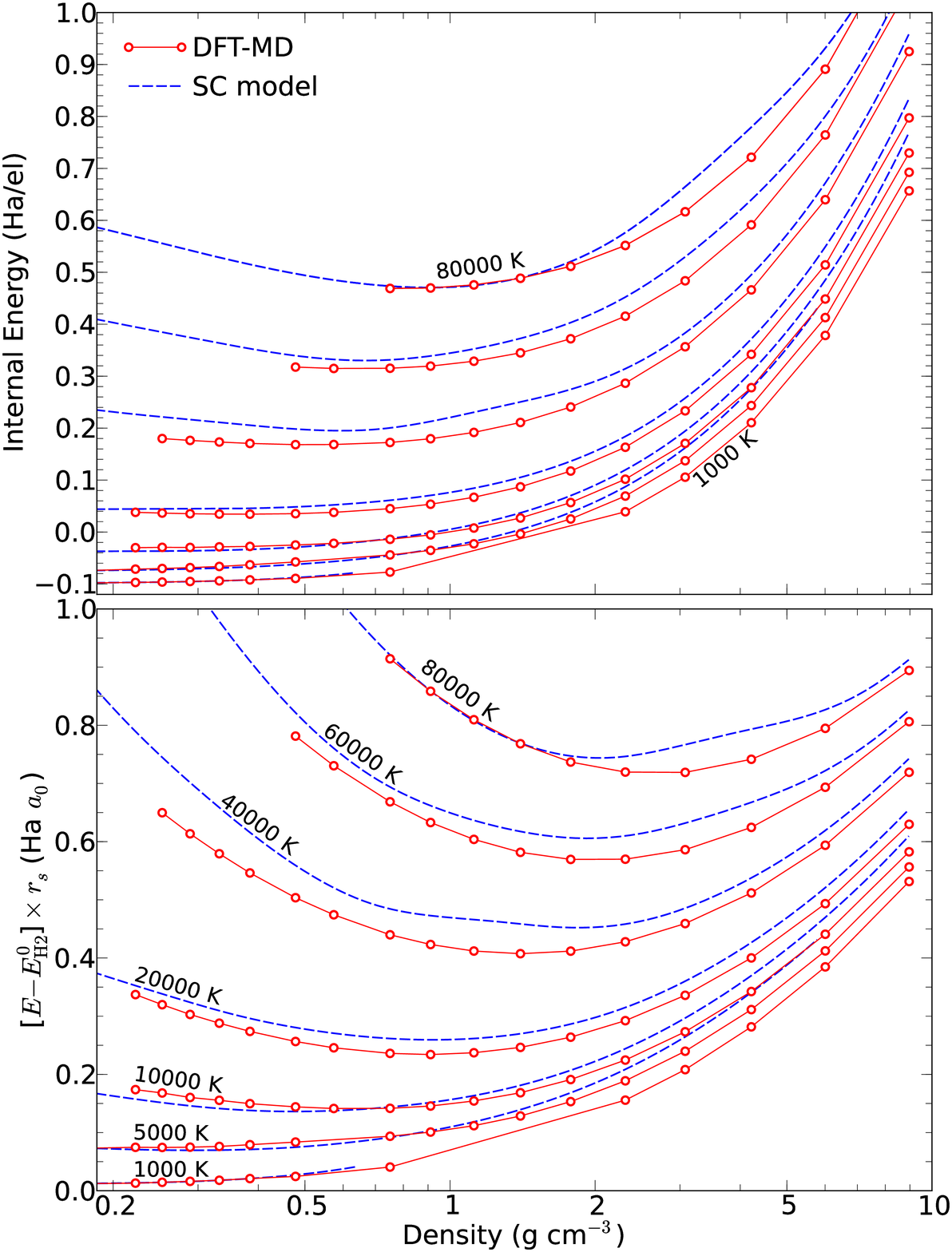}
\caption{Internal energy per electron as function of density for seven
  different temperatures. Predictions from DFT-MD simulations and from
  the analytical SC EOS model are compared. \label{EOSErho}}
\end{figure}

We report the computed equation of state in the form of a table, a series
of figures, and in analytical form as two-dimensional fit of the free
energy. In table~\ref{EOS}, we provide the thermodynamic functions
that directly follow from analysis of the DFT-MD trajectories.
The pressure and internal energy were computed
for 391 different density-temperature points (see Fig.~\ref{EOSTrs}).
The 1-$\sigma$ error
bars correspond to statistical uncertainty that arises from the finite
length of the MD simulations. For 131 points in table~\ref{EOS}, the
thermodynamic integration was performed with five $\lambda$ points and
the free energy and entropy are reported in addition. Only counting
the production runs that led to results in table 1, the total CPU time
consumed for this project amounted to 850$\,$000 core-hours on Intel
Nehalem processors. This is equivalent to using 100 cores for an
entire year, which is a considerable amount of computer time by
today's standards but will certainly become available to everyone in
the near future as computers with more and more cores are assembled.

\begin{figure}
\epsscale{1.0}
\plotone{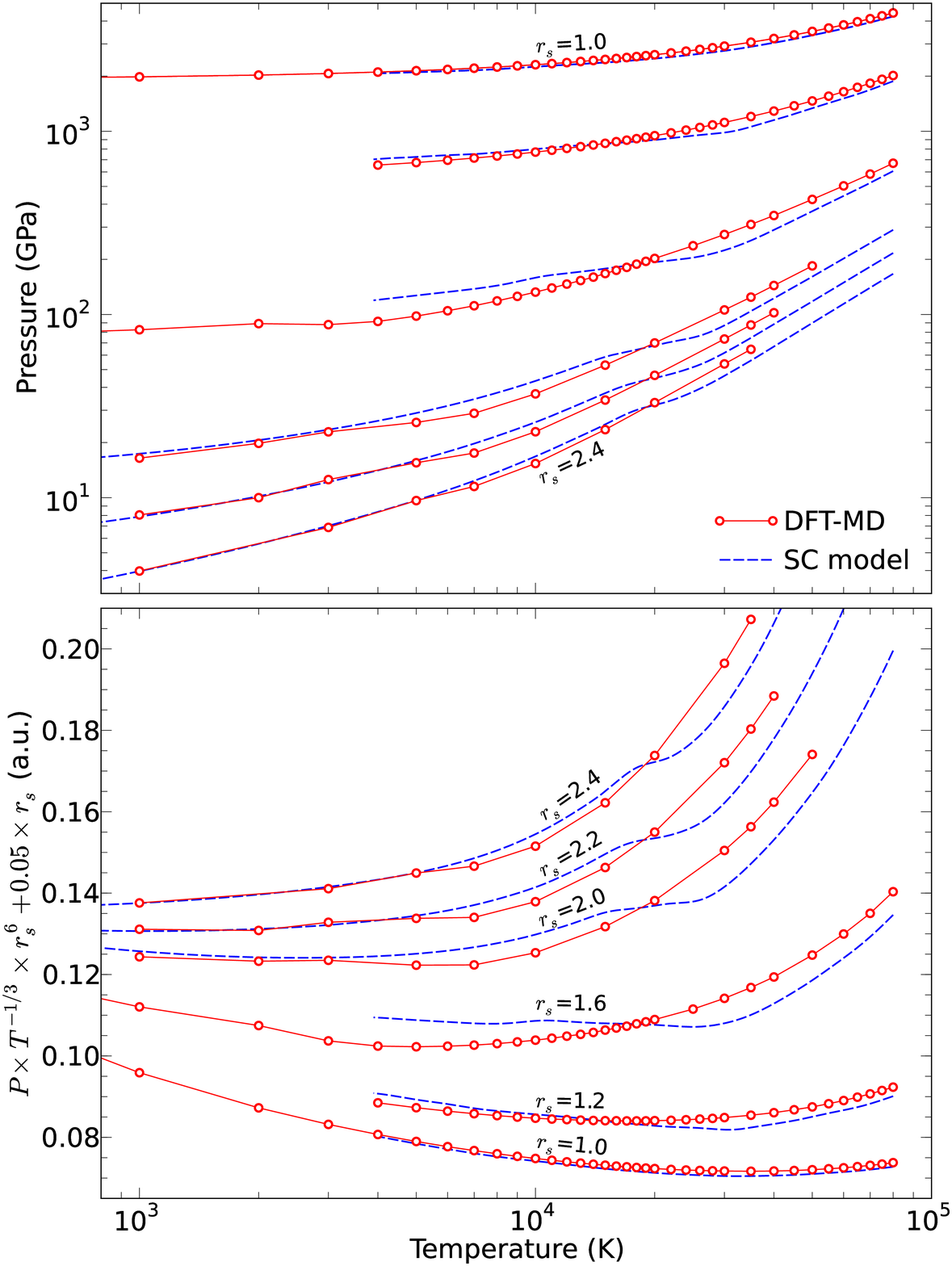}
\caption{Pressure isochores computed with DFT-MD simulations are compared with the analytical SC model. \label{EOSPT}}
\end{figure}

\begin{figure}
\epsscale{1.0}
\plotone{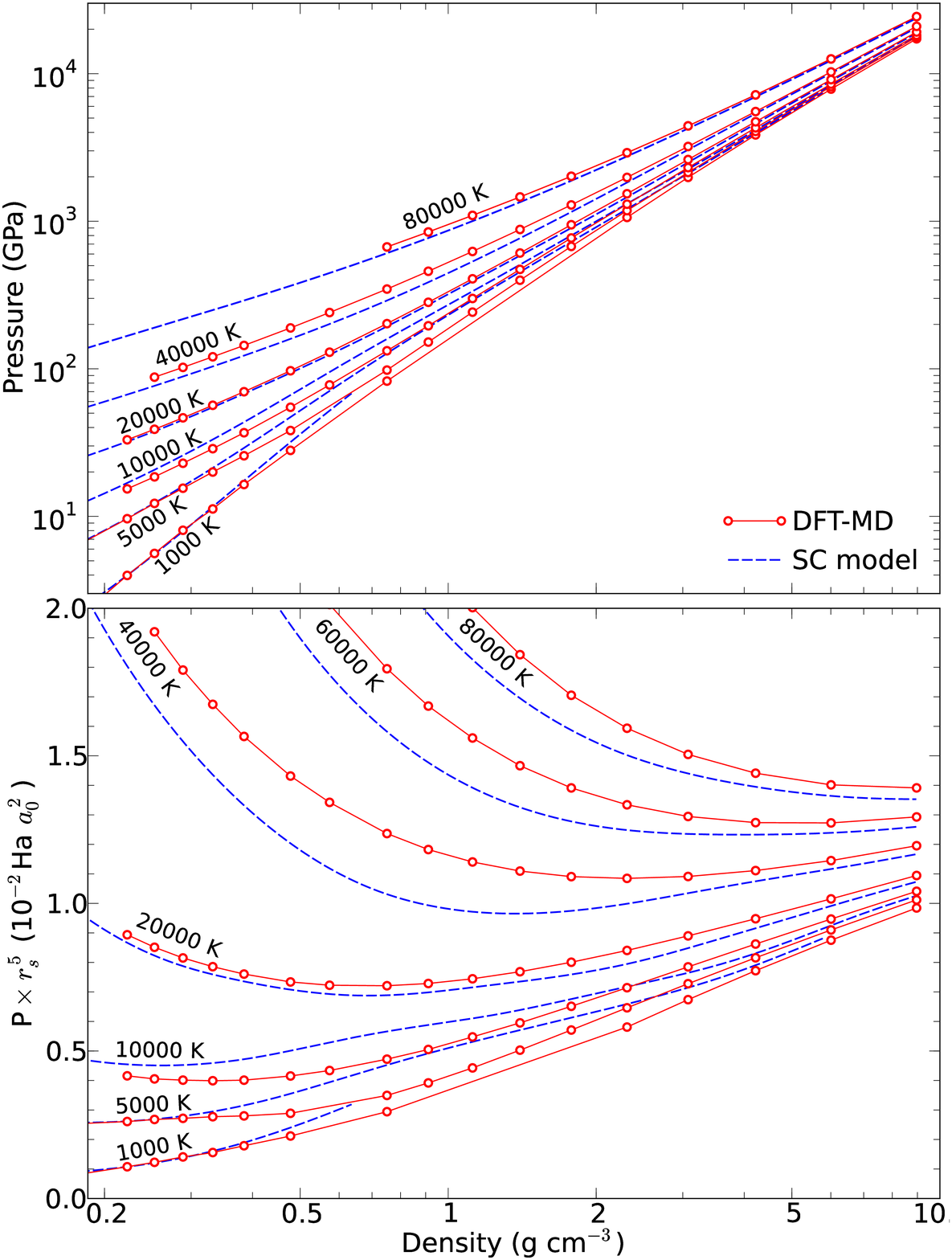}
\caption{Pressure isotherms computed with DFT-MD simulations are compared with the analytical SC model. \label{EOSPrho}}
\end{figure}

\begin{figure}
\epsscale{1.0}
\plotone{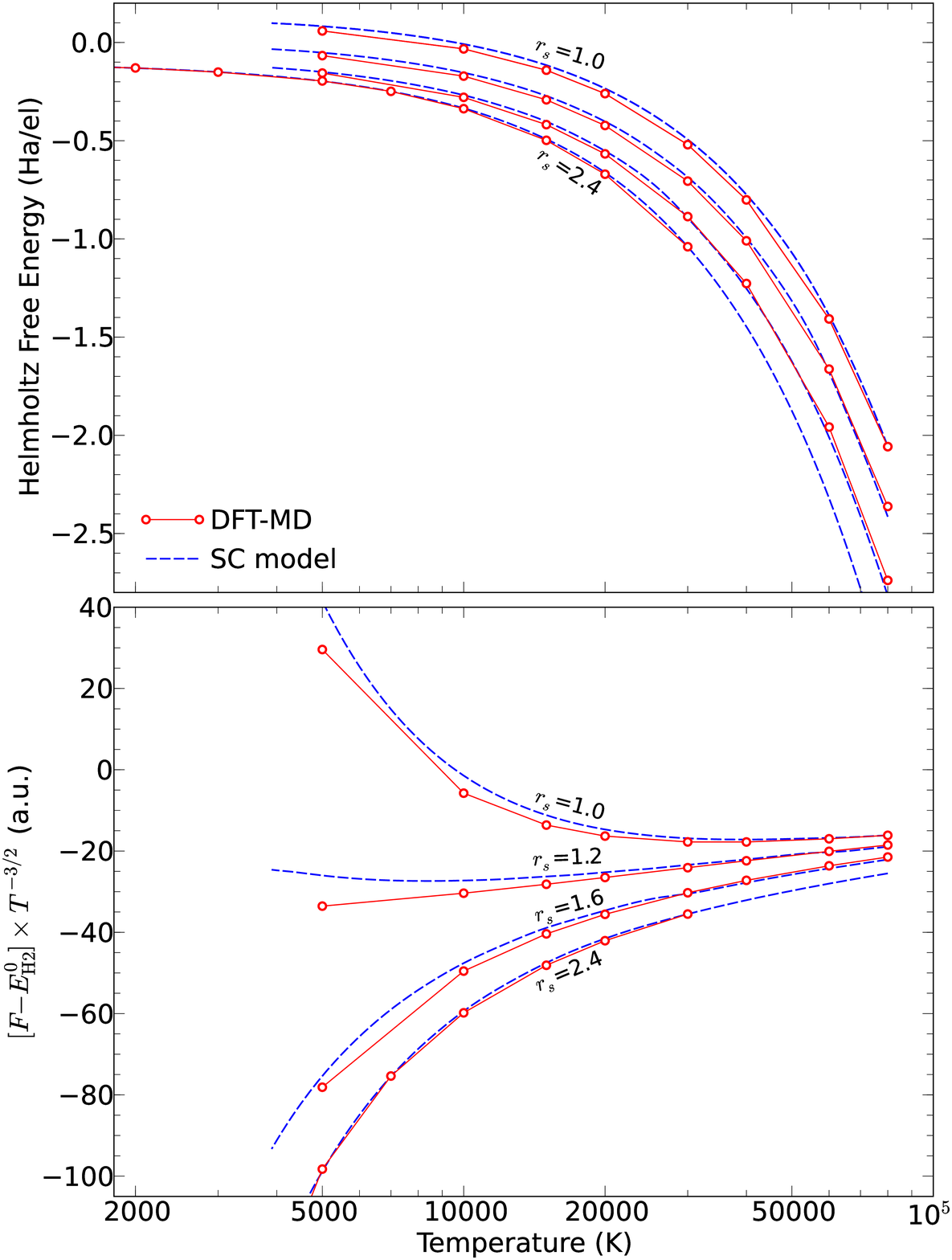}
\caption{Helmholtz free energy per electron at constant density computed with DFT-MD simulations are compared with the analytical SC model. \label{EOSFT}}
\end{figure}

\begin{figure}
\epsscale{1.0}
\plotone{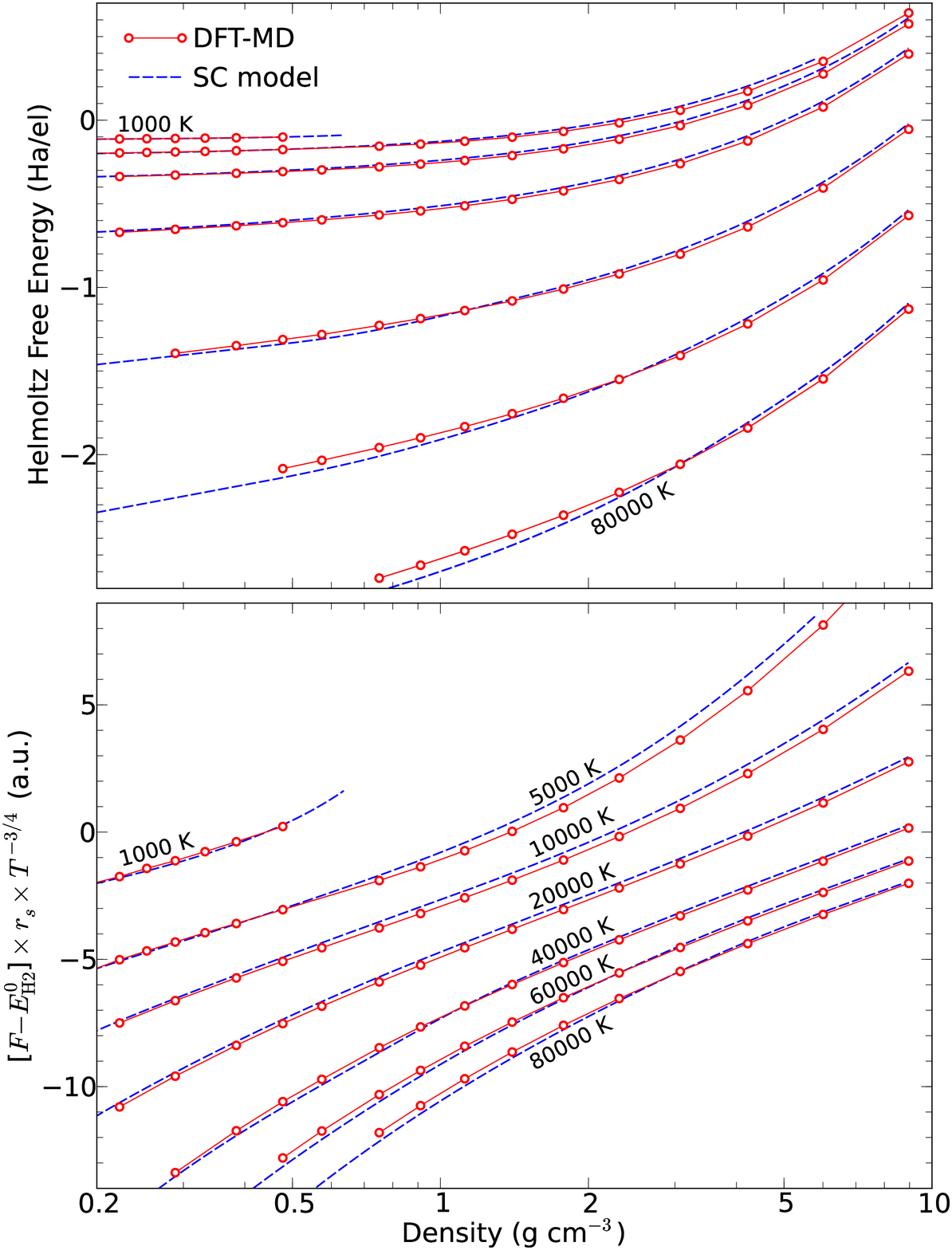}
\caption{Helmholtz free energy per electron at constant temperature computed with DFT-MD simulations are compared with the analytical SC model. \label{EOSFrho}}
\end{figure}

\begin{figure}
\epsscale{1.0}
\plotone{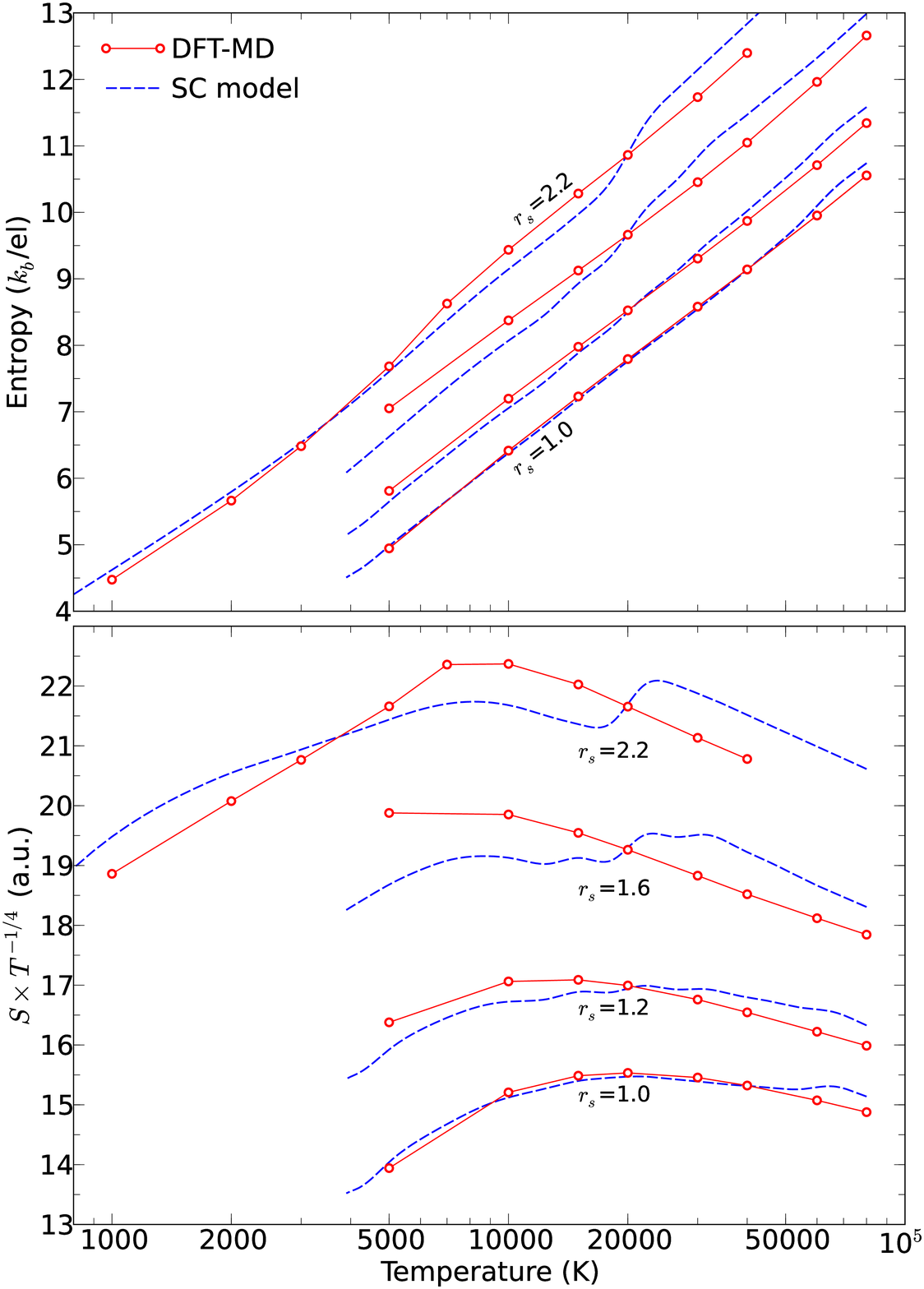}
\caption{Entropy per electron at constant density computed with DFT-MD simulations are compared with the analytical SC model. \label{EOSST}}
\end{figure}

\begin{figure}
\epsscale{1.0}
\plotone{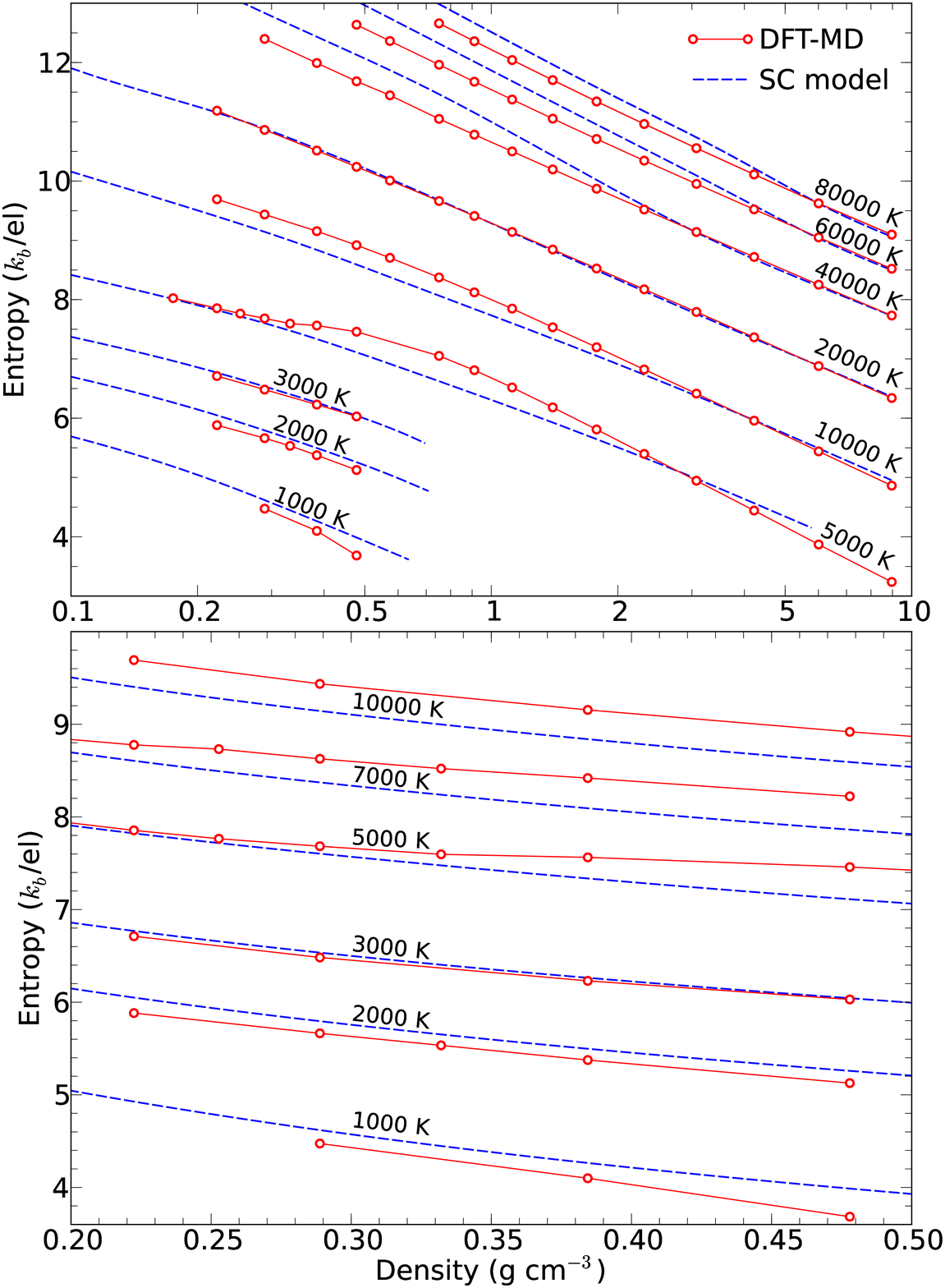}
\caption{Entropy per electron at constant temperature computed with
  DFT-MD simulations are compared with the analytical SC model. The
  upper panel shows results over a wide temperature-density range
  while the lower panel zooms in on the low density regime where hydrogen
  occurs in molecular form when the temperature is below 5000$\,$K. \label{EOSSrho}}
\end{figure}

In figures~\ref{EOSET}, \ref{EOSErho}, \ref{EOSPT}, \ref{EOSPrho},
\ref{EOSFT}, \ref{EOSFrho}, \ref{EOSST}, and \ref{EOSSrho}, we plot
the internal energy, pressure, Helmholtz free energy, and entropy
respectively as a function of temperature and density. Every circle
corresponds to a particular DFT-MD simulation listed in
table~\ref{EOS}, without any interpolation being performed. The dashed
lines are the results of the most common version of the analytical SC EOS
model where the different thermodynamic functions have been smoothly
interpolated across the molecular-to-metallic transition in hydrogen.

To accommodate the wide parameter range of our simulations, we plot
the different thermodynamic functions on logarithmic scale. Since
these functions depend strongly on density and temperature, we added a
second panel where we removed most of this dependence by introducing a
scale factor equal to $r_s$ or $T$ raised to some power. Here $r_s$ is the
Wigner-Seitz radius that specifies the density of system according to
$\frac{4 \pi}{3} r_s^3=V/N_e = n^{-1}$, while $n$ is the number of
electrons, $N_e=(N_{\rm H} + 2 N_{\rm He})$, per unit volume $V$. The
mass density is given by $\rho = n (N_{\rm H} m_{\rm H} + N_{\rm He}
m_{\rm He})/(N_{\rm H} + 2 N_{\rm He})$ where $m_{\rm H}$ and $m_{\rm
  He}$ are the masses of the hydrogen and helium atoms.

The rescaling of the ordinate makes it easier to identify the
deviations from the SC model while our simulation results can still
be reproduced easily. The ordinates are plotted in atomic units.
Lengths including $r_s$ are given in Bohr radii ($a_0$=5.29177209$\times
10^{-11}\,$m), energies in Hartrees (4.35974380$\times 10^{-18}\,$J)
per electron (el.), and entropies are specified in units of $k_b$ per
electron, where $k_b$ is Boltzmann's constant.

Figure~\ref{EOSET} shows a comparison between the internal energies
from DFT-MD simulations with the predictions of the SC EOS model. For
a low density of $r_s=2.4$, excellent agreement is found for a
temperature range from 1000 to 20$\,$000$\,$K. Hydrogen gradually
changes from a molecular state to an atomic state in this temperature
interval and, from the good agreement, one may conclude that the
thermally activated dissociation of molecules is well described in the SC
model. However, 20$\,$000$\,$K, the SC model predicts an strong and
artificial increase in the internal energy that is the result of an
inaccurate description of electronic excitations. This deviation was
first identified by~\citet{MC01} when predictions from the SC
model were compared with PIMC simulations.  Figure~\ref{EOSErho} shows
that this deviation is present at 20$\,$000$\,$K for whole density
range under consideration and extends to much higher temperatures
also.

Figure~\ref{EOSET} shows that the favorable agreement between DFT-MD
results and SC predictions below 20$\,$000$\,$K continues to hold up
to a density of $r_s=1.6$. When the internal energy is compared for a
higher density of $r_s=1.0$ or 1.2 where hydrogen is metallic, one
finds that DFT-MD results and SC predictions are offset by a nearly
constant amount.

The internal energy curves of $r_s=1.6$ and 2.4 appear to cross over
in Fig.~\ref{EOSET} at a temperature of 27$\,$000$\,$K, which is
consistently predicted by DFT-MD results and the SC model.
Figure~\ref{EOSErho} shows that this is simply a consequence of
internal energy exhibiting a minimum when plotted at constant
temperature as function of density. At high density, the internal
energy sharply rises because of Pauli exclusion effects between the
electrons.  In the low density limit, the internal energy rises also
because the ionization fraction increases as a result of the increased
gain in entropy that is associated with electrons becoming free
particles.

In Fig.~\ref{EOSPT}, we compare the pressure predicted from DFT-MD
simulation with the SC model. At a high density of $r_s=1.0$ where the
hydrogen-helium mixture is metallic, we find fairly good agreement
over the entire temperature range. This implies that the deviation
that we identified for the internal energy in this regime, varies
slowly with density and does not significantly affect the pressure in
the SC model.

At a low density of $r_s=2.2$ and 2.4, we found good agreement up to a
temperature of 5000$\,$K. At this temperature, we see a small decrease
in slope in the DFT-MD data that is missing in the predictions of the
SC model. We attribute this slope change to the dissociation of
molecules in the DFT-MD simulations. At 20$\,$000$\,$K, the SC model
predicts a significant decrease in slope which is not present in the
DFT-MD data. This slope change in the SC predictions can again be
attributed to an inaccurate description of ionization, which leads to
deviations over the whole density range under consideration
(Fig.~\ref{EOSPrho}). At an intermediate density of $r_s=1.6$ close to
the molecular-to-metallic transition, we find that the SC model
overestimates the pressure up to about 20$\,$000$\,$K and
underestimates for higher temperatures. The deviations around 100 GPa,
5000$\,$K, and $r_s=1.6$ (0.75 g$\,$cm$^{-3}$) are of particular
significance. The DFT-MD simulations predict pressures that are much
lower than those of the SC model. This leads a significant departure
in the resulting adiabats. Its implication for the interiors of giant
planets will later be further analyzed.

In figures~\ref{EOSFT} and \ref{EOSFrho}, the Helmholtz free
energy from DFT-MD simulations and the SC model are compared. In
general the agreement appears to be much better than for other
thermodynamic functions that are derivatives of it. Still, one finds
that the SC model overestimates the free energy in the metallic regime,
mirroring the deviations that we have discussed for the internal
energy.

In figures~\ref{EOSST}, the entropies at different densities are
compared as a function of temperature. At a very high density of
$r_s=1.0$, very good agreement between DFT-MD results and the SC model
is found up to 50$\,$000$\,$K. For lower densities, the SC model
predicts a sharp entropy increase at 20$\,$000$\,$K, which is again a
result of the treatment of ionization effects. This trend is not
confirmed by the DFT-MD simulations. One also finds significant
deviations at lower temperatures, in particular around $r_s$=1.6. Even
at a relatively low density of $r_s=2.2$, the agreement is not
perfect. From 4000 to 20$\,$000$\,$K, the SC model underestimates the
entropy and it overestimates the entropy for lower temperatures.
Figure~\ref{EOSSrho} shows that such deviations persist over a wider
density range. In principle, one expects a non- or weakly interacting
gas of hydrogen molecules and helium atoms to be perfectly described
by the SC model. However, the density that we can efficiently study
with DFT-MD simulations does not yet appear to be low enough for the
deviations to decay to zero.

\begin{figure}
\epsscale{1.0}
\plotone{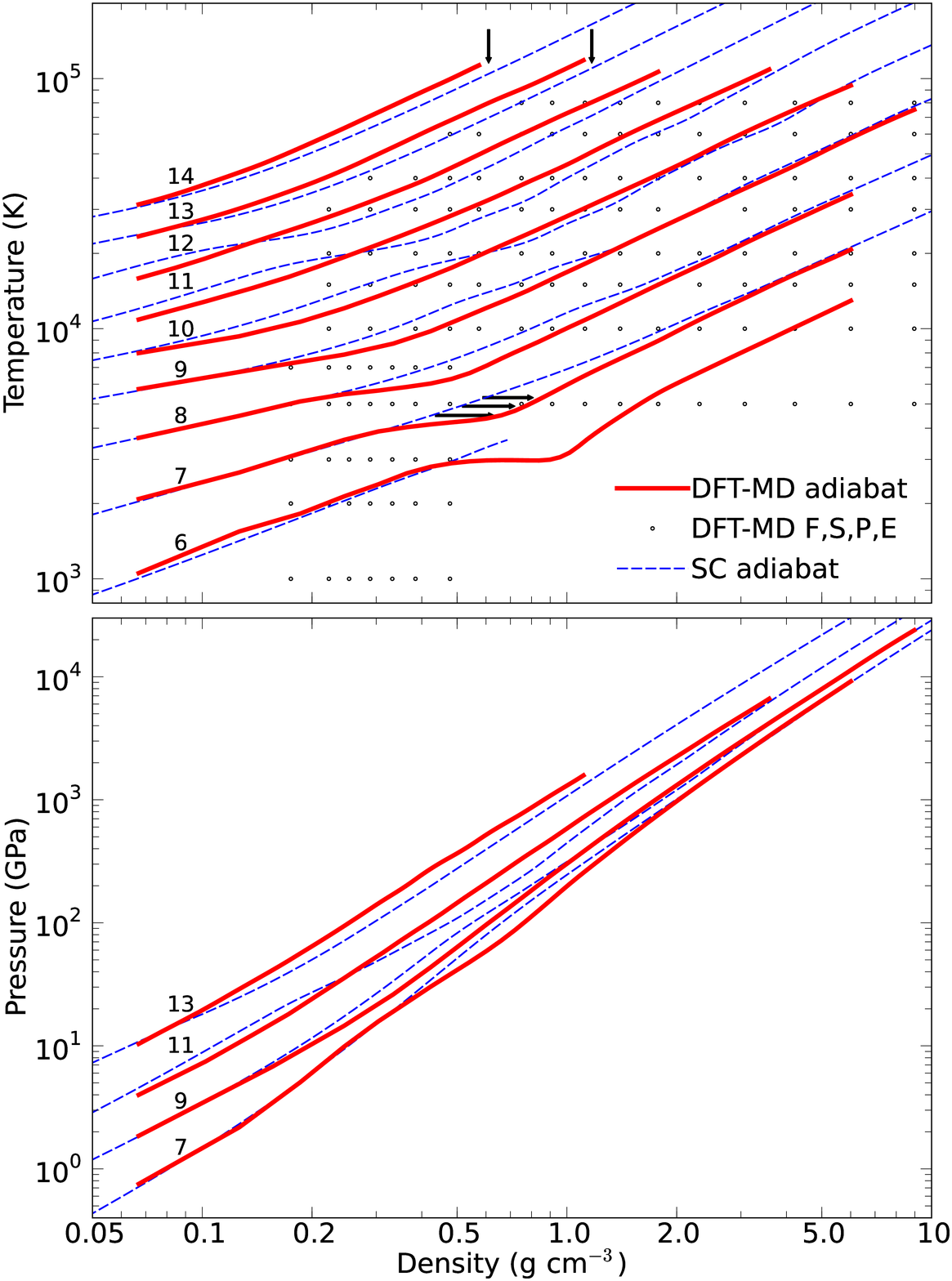}
\caption{Adiabats derived from DFT-MD simulations are compared with
  the SC model. The labels denote the entropy in units of $k_b$ per
  electron. The circles indicate parameters where entropy and free
  energy have been calculated in addition to the pressure and internal
  energy. The horizontal arrows label conditions where the deviation
  from the SC model are large and important for the interiors of
  Saturn and Jupiter. The vertical arrows indicate deviations are high
  temperature where the SC model does not treat electronic excitations
  accurately.\label{ad}}
\end{figure}

\section{Free Energy Fit for the Equation of State}

We fitted our {\it ab initio} results for $P$, $E$, $F$, and $S$ in
table~\ref{EOS} with a two-dimensional spline function that represents
the Helmholtz free energy in terms of temperature, $T$, and electron density,
$n=N_e/V$. By construction,
this fit is thermodynamically consistent. We employ the same
functional form that we used to represent the free energy of hot,
dense helium in~\citep{Mi09}, except the splines here are functions of
$n$ rather than $\log(n)$. Table~\ref{EOS_fit} provides the free
energy as well as the required derivatives on a number of $(n,T)$ knot
points.  Atomic units are used throughout.

To evaluate the fit for $(n^*,T^*)$, we first construct a separate
one-dimensional cubic spline function, $F_n(T)$, for every density on
a grid ranging from $r_s$=3.581 to 0.536
(0.0670$-$20.0$\,$g$\,$cm$^{-3}$). At every density, the free energy
is given on a number of temperature knots and its first derivative,
$\left.\frac{\partial F}{\partial T}\right|_n$, is specified for the
highest and lowest temperatures. We construct a similar
one-dimensional spline function that represents $\left.\frac{\partial
    F}{\partial n}\right|_T(T)$ at the smallest and largest density.
We then evaluate all these splines functions at $T^*$ and construct a
one-dimensional spline function $F_{T^*}(n)$ from the free energy
values and its first derivatives at the boundaries. This provides us
not only with a straightforward way to obtain the free energy at every
$(n,T)$ point but we can also derive the pressure and entropy by
taking analytical derivatives,
\begin{equation}
P=n^2\left.\frac{\partial F}{\partial n}\right|_T
\;\;{\rm and}\;\;
S=-\left.\frac{\partial F}{\partial T}\right|_n
\;\;.
\end{equation}
The internal energy and Gibbs free energy then follow from $E=F+TS$
and $G=F+PV$.

When we constructed this fit, we made sure every EOS point in
table~\ref{EOS} is well reproduced. We extended the domain of the fit
a bit beyond the range of the DFT-MD data. This leads to a smoother
representation of the data in the interior of the domain and also
allows us to gradually approach the SC EOS in the limit of low
density. As figure~\ref{ad} shows, we were able to smoothly match onto
the SC adiabats for entropy values from 6 to 10 and again for 13 and 14
$k_b$/el. A disagreement remains for S=11 and 12 $k_b$/el. but the SC
EOS is not thermodynamically consistent in the regime of 10$\,$000$\,$
to 20$\,$000$\,$K and no attempt was made to reproduce those adiabats.
So far, only the one exoplanet HAT-P-32b ~\citep{Hartman2011} appears to
have an internal entropy in excess of 11 $k_b$/el.; see Figure~\ref{RvsM}.

We also find a significant disagreement in the high density limit
between our DFT-MD adiabats and the predictions of the SC model.
Starting with entropy values of $S=10\,k_b$/el., the DFT-MD results
predict the adiabats reach states of higher temperatures and higher
pressure for a given density.  

The most significant result of Fig.~\ref{ad} is the deviations along
the $S$=7 $k_b$/el. adiabat. The DFT-MD simulations predict a decrease
in slope of the adiabat exactly where the hydrogen molecules
dissociate~\citep{MHVTB}.  Since the SC model interpolates between
separate atomic/metallic and molecular thermodynamic descriptions, it
has no predictive power in the regime of pressure dissociation where
all the different species interact strongly.

\section{Giant Planet Interiors}

\begin{figure}
\epsscale{1.0}
\plotone{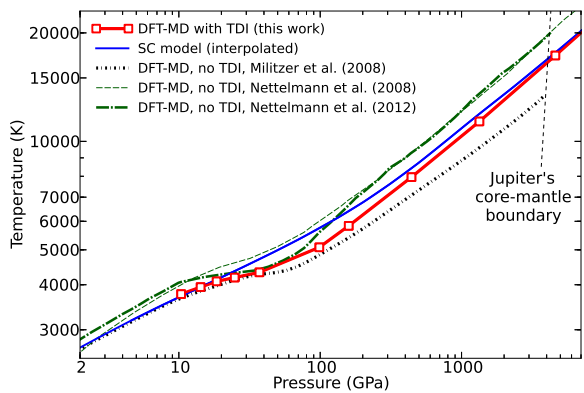}
\caption{Adiabats from different calculations for Jupiter's interior.\label{jupiter}}
\end{figure}

In Fig.~\ref{jupiter}, we compare different predictions for the
adiabat in Jupiter's interior. Similar to~\citet{MHVTB}, the
calculations of the entropy by~\citet{NHKFRB,Nettelmann_2012} relied
solely on the $P$ and $E$ from DFT-MD simulations. Since no TDI was
employed, the entropy was determined indirectly from thermodynamic
relationships and an integration over a large path through
density-temperature space.  Here one faces two challenges. Since
the integration can only yield the entropy difference between two
$\rho$-$T$ points, one needs to find a starting point for the
integration where DFT-MD simulations work and the entropy is known
reliably through other means. Secondly, one needs to determine $P$
and $E$ on a very fine grid in $\rho$-$T$ space, so that
integration errors do not accumulate. Since both challenges are
difficult to meet, we instead adopted the more reliable TDI technique
that we extended to molecular systems in~\citet{Militzer2013}.

Figure~\ref{jupiter} shows that there exist some discrepancies between our
TDI calculations and the~\citet{NHKFRB,Nettelmann_2012} results in the
low density as well as the high density limit. In the molecular regime
at 10 GPa, Nettelmann et al.'s temperatures for Jupiter's adiabat are 8\%
higher than our TDI calculations predict. In the metallic regime at
4000 GPa, their results are 19\% higher than our TDI predictions.
This implies that there exist discrepancies in the low and high
density limits before any {\it ab initio} $P$ and $E$ data points are
entered into the calculation of the adiabats
by~\citet{NHKFRB,Nettelmann_2012}.

Adiabats based on the ~\citet{Nettelmann_2012} work now show a
pronounced flattening in the regime of molecular dissociation from
15-40 GPa that was not present in the ~\citet{NHKFRB} calculations.
The pressure range similar to our {\it ab initio} results but the
magnitude is higher than we predict based on TDI. From 40 to 200 GPa,
the~\citet{Nettelmann_2012} calculations predict a steep rise in
temperature for the adiabats. This is not consistent with our TDI
calculations and implies that in models by ~\citet{Nettelmann_2012}
most of Jupiter's mass is at 19\% higher temperature than we predict
based on our TDI calculations. In the low and high density limits, our
adiabats are instead in relatively good agreement with the SC model.

\section{Mass-Radius Relationships}

\begin{figure}
\epsscale{1.0}
\plotone{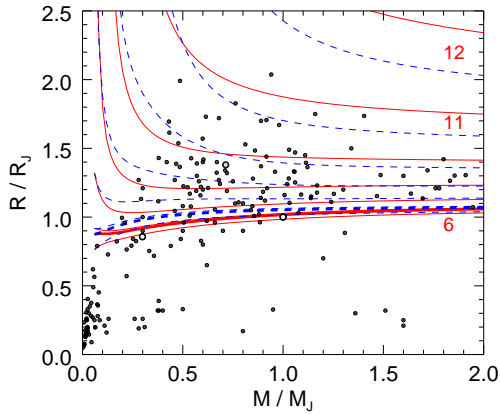}
\caption{Radius $R$ (in units of $R_J = 70000$ km vs. mass $M$ (in
  units of $M_J =$ Jupiter's mass). The solid data points are
  measurements of transiting exoplanets. The curves show the $R(M)$
  relation predicted from two EOS calculations (solid curves are for
  DFT-MD simulations, dashed curves are for the SC model) for fixed
  entropies of $S=$ 6, 6.8 (Saturn; heavy curve), 7 (Jupiter), 8, 9,
  10, 11, and 12 $k_b$/el. The three open data points denote Saturn,
  HD 209458b, and Jupiter from left to right. \label{RvsM}}
\end{figure}

The vexing problem of radius anomalies of transiting giant planets
~\citep{Burrows_2007} has continued with the addition of more objects
~\citep{Laughlin_2011}. Figure~\ref{RvsM} shows measurements posted in
the online Exoplanet Encyclopaedia~\citep{Schneider_2011} as of late
2012 and different theoretical curves that we will discuss below.
Briefly, the problem arises from a significant population of
exoplanets that have radii too large to be explained by thermal
distention from retained primordial heat, and there is a further
population with radii well below those expected for primarily H-He
composition even at zero temperature.  Any point that falls below the
$S=6\,k_b$/el. curve can be explained by invoking the presence of a
rocky core and/or admixture of heavy elements, which reduces the
radius for given mass~\citep{MF_2011}.  But it is not so simple to
classify the population of anomalously distended giant exoplanets, for
the degree of distention depends on such factors as the planet's age and
degree of irradiation from the host star~\citep{FN_2010}, and possible
additional heating mechanisms such as ohmic
dissipation~\citep{Bat_2011}.

In order to most clearly exhibit the differences in predicted radius
between the DFT-MD simulations and analytical SC model, we model a
planet of mass $M$ as a H-He object of constant entropy $S$ and fixed
helium fraction of $Y=0.245$ with neither a rocky core nor heavy element
component in the gas envelope. Since we do not have DFT-MD simulation
data at very low densities, we switch back to the SC model below
0.0670$\,$g$\,$cm$^{-3}$, the lowest density of the free energy fit to
our DFT-MD data.

\begin{figure}
\epsscale{1.0}
\plotone{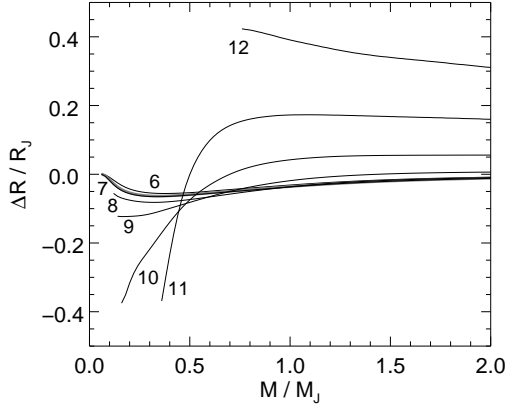}
\caption{The ordinate is the radius difference
$\Delta R = R_{\rm DFT-MD} - R_{\rm SC}$ for a given adiabat.  The heavy 
shaded curve is for a Saturn adiabat with $S =$ 6.8.  At low masses
(below about 0.1 $M_J$ for low entropies) the difference goes to
zero because there are no DFT-MD data at low densities. \label{dRvsM}}
\end{figure}

As is well known~\citep{ChandraSS}, formally such an object
has a precisely defined radius where the temperature $T$, mass density
$\rho$, and pressure $P$ simultaneously go to zero.  In a real object,
the ideal-gas outer layers cannot persist in such an isentropic state
and instead the temperature reaches a finite limit set by the
effective temperature for the radiation balance in the outer layers.
The radius as measured by transit observations also depends on sources
of slant opacity in these outer layers.  As explored
by~\citet{Burrows_2007}, the value of such a radius can vary by
several $0.1\,R_J$, depending on the atmospheric model.  Adjusting the
parameters controlling the atmosphere can move a model closer to
agreement with objects of inflated radii, but sometimes a mismatch
remains.  A major result of the present paper is that differences in
the EOS alone can also lead to radius changes of several $0.1\,R_J$.
Because we consider only strict-adiabatic models, the deviations that
we point out are entirely due to differences in the EOS at high
pressure. In Figure~\ref{dRvsM}, we exhibit these differences for
various entropies. For entropy values up to 9 $k_b$/el., the DFT-MD
calculations consistently predict smaller planet radii than the SC
model, which is a direct consequence of the density enhancement on the
adiabats around 100 GPa illustrated in Figs.~\ref{ad} and
\ref{jupiter}. Figure~\ref{ad} also shows that the DFT-MD and SC
adiabats for 10, 11, and 12 $k_b$/el. cross over in the density range
from 0.15 to 0.7$\,$g$\,$cm$^{-3}$. This is the reason why the DFT-MD
calculation predict larger planet radii than the SC model for massive
planets with $M > 0.5 M_J$ but significantly smaller radii for light
planets. The deviations between the DFT-MD and SC predictions in
Fig.~\ref{dRvsM} reach values up to approximately 0.4 Jupiter radii.

\begin{figure}
\epsscale{1.0}
\plotone{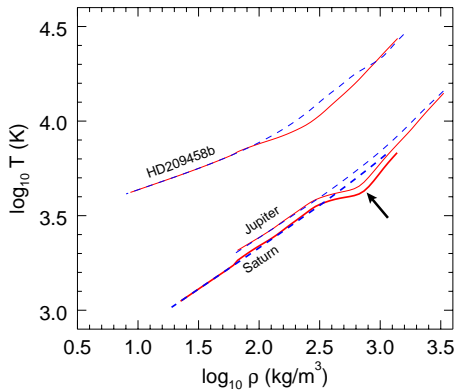}
\caption{Temperature-density profile for the interior of three specific
  objects.  Solid curves are for DFT-MD and dashed curves are for SC.
  The enhanced density in the DFT-MD model (see arrow) at pressures near $100$ GPa
  produces a slightly smaller radius for both Jupiter and Saturn.
  Below a density of 0.0670$\,$g$\,$cm$^{-3}$, the SC model is used
  for both calculations. \label{T-rho}}
\end{figure}

\begin{figure}
\epsscale{1.0}
\plotone{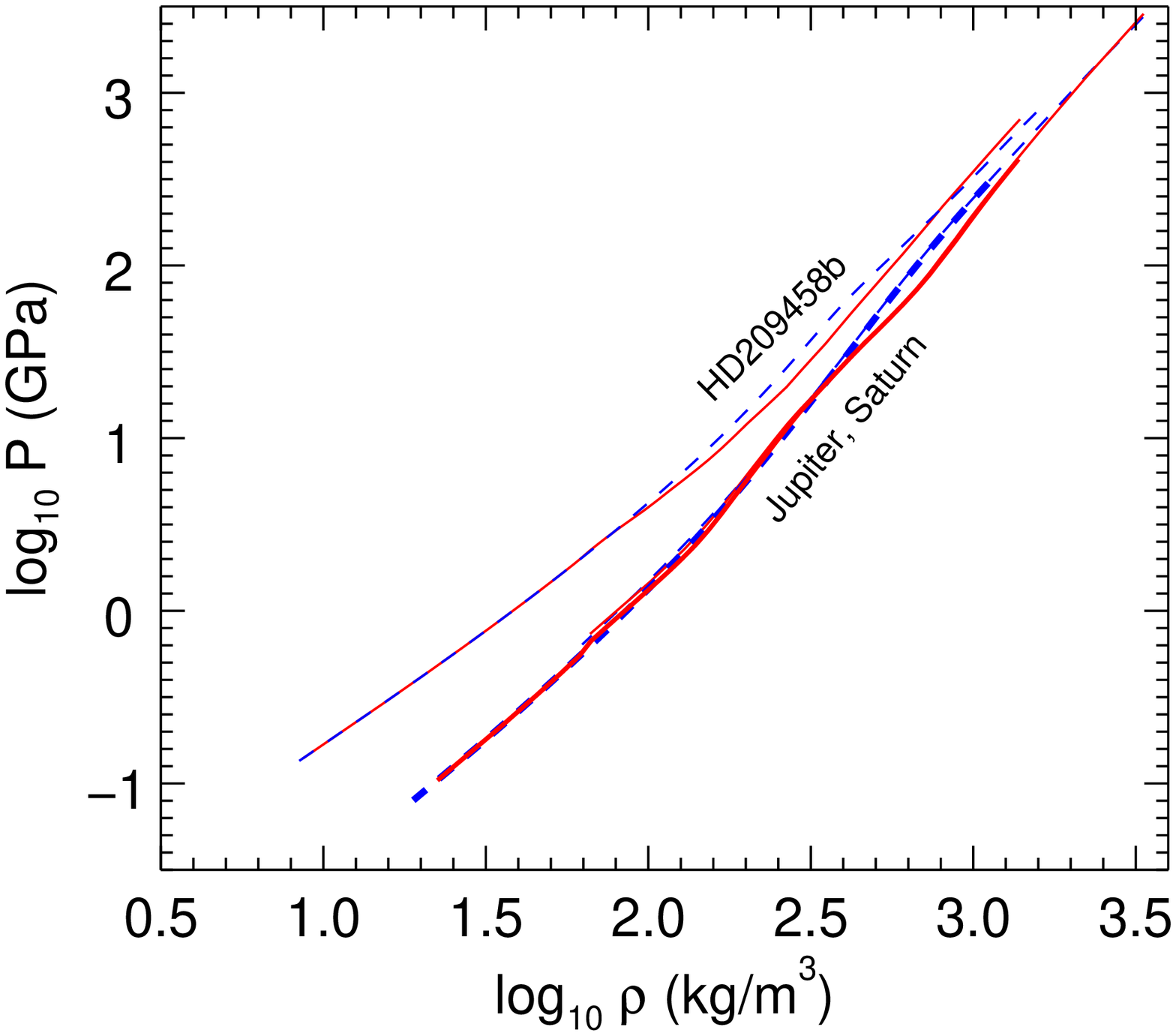}
\caption{Pressure-density profile for the planetary interiors shown in
  Fig.~\ref{T-rho}. \label{P-rho}}
\end{figure}

The exoplanet HD 209458b (middle open data point in Figure~\ref{RvsM})
fortuitously falls near an entropy $S \approx 9.5$ $k_b$/el. and mass $M \approx
0.7 M_J$ where $\Delta R \approx 0$. Nevertheless, the interior
$T$-$\rho$ and $P$-$\rho$ profiles in figures~\ref{T-rho} and
\ref{P-rho} differ significantly between the DFT-MD and SC EOSs.
These figures also compare interior profiles for simplified (pure H-He
mixtures on an adiabat) models of Jupiter and Saturn.

\begin{figure}
\epsscale{1.0}
\plotone{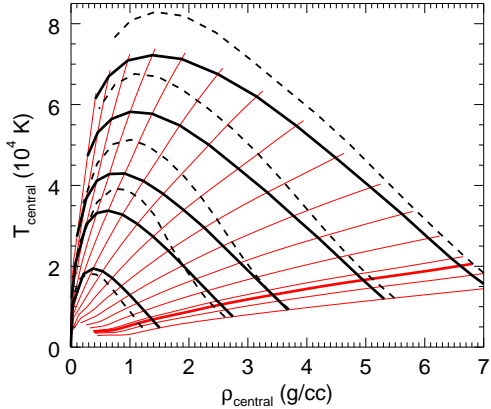}
\caption{The values of $T$ and $\rho$ at the center of a planets
  according to the DFT-MD (heavy solid lines) and SC (dashed lines)
  EOSs for five different planet masses corresponding to 0.3 $M_J$
  (Saturn, lowest curves), 0.7 $M_J$ (HD 209458b), 1.0 $M_J$
  (Jupiter), 1.5 $M_J$, and 2.0 $M_J$ (top curves).  The thin solid
  lines refer to various DFT-MD adiabats with entropy values from 6 (lowest
  curve) to 12 (steepest curve) in steps of 0.5 $k_b$/el.  The heavy
  solid line is for Saturn's entropy of 6.8.
  \label{Tcrhoc}}
\end{figure}

Figure~\ref{Tcrhoc} shows differences in evolutionary behavior of our
simplified planetary models.  This figure plots the value of the
central temperature, $T_{\rm central}$, vs. the central density,
$\rho_{\rm central}$, for a range of adiabats and masses. During the
evolution of a planet of constant mass, its central density increases
monotonically while its central temperature exhibits a maximum.
% {\bf  (Please check.)}
During the initial contraction, the temperature in
the center increases at first as the material is subjected to
increasing pressure. When a degenerate interior state is reached, the
contraction ceases and the whole planet starts to cool. According to
DFT-MD simulations the maximum temperature reached is up to 10$\,$000$\,$K
lower than
predicted by the SC model. This deviation may have consequences for
the evolution of cores in giant planets that remain to be explored.

\section{Conclusions}

This paper provides an equation state table for hydrogen-helium
mixtures in giant planet interiors that was derived from {\it ab
  initio} computer simulations. The combination with an efficient
thermodynamic integration technique enabled us to calculate the
entropy and free energy directly, in addition to pressure and internal
energy that follow from standard simulations. 

Our complete EOS table with 391 density-temperature points as well as
a thermodynamially consistent free energy fit is included in this
publication so that our EOS can be easily incorporated in future
models for giant planet interiors.

We have identified significant deviations for the Saumon and Chabrier
EOS models. The new DFT-MD EOS causes low-entropy giant-planet models
($S \le 8\, k_b$/el.) to shrink in comparison to SC models by up to
0.08 Jupiter radii. But for hot giant planets with mass exceeding
$0.5\,M_J$ and with interior entropy values in the range from
10$-$12$\,k_b$/el., the DFT-MD simulations predict significantly
larger radii. The correction to the SC model reaches 0.4 Jupiter radii
for the hottest planets. Thus, the revision suggests that some of the most
inflated giant exoplanets are at lower entropies than was previously
inferred.  Our revision could ameliorate the ``inflated giant exoplanet''
discrepancy to some extent but perhaps not for HD209458b.
The matter is to be
revisited with detailed evolutionary calculations based on our revised
EOS.

\acknowledgments

This work has been supported by NASA and NSF. Computational resources
at NCCS were used.

%% {\it Facilities:} \facility{Nickel}, \facility{HST (STIS)}, \facility{CXO (ASIS)}.



\appendix




%\bibliography{hhe}{}
%\bibliographystyle{apj}

\clearpage

%% the file, uncomment the \plotone, \plottwo, and
%% If you need a layout that cannot be achieved with \plotone or
%% \plottwo, you can invoke the graphicx package directly with the
%% \includegraphics command or use \plotfiddle. For more information,
%% please see the tutorial on "Using Electronic Art with AASTeX" in the
%% documentation section at the AASTeX Web site,
%% http://www.journals.uchicago.edu/AAS/AASTeX.
%%
%% The examples below also include sample markup for submission of
%% supplemental electronic materials. As always, be sure to check
%% the instructions to authors for the journal you are submitting to
%% for specific submissions guidelines as they vary from

%% This example uses \plotone to include an EPS file scaled to
%% 80% of its natural size with \epsscale. Its caption
%% has been written to indicate that additional figure parts will be
%% available in the electronic journal.

%\begin{figure}
%\epsscale{.80}
%\plotone{EOSTrs12}
%\caption{Derived spectra for 3C138 \citep[see][]{heiles03}. Plots for all sources are available
%in the electronic edition of {\it The Astrophysical Journal}.\label{fig1}}
%\end{figure}

\clearpage

\clearpage

\begin{deluxetable}{cccrlll}
\tabletypesize{\scriptsize}
\tablecolumns{6}
\tablewidth{0pc}
\tablecaption{Equation of state derived from DFT-MD simulations.\label{EOS}}
\tablehead{
  \colhead{$r_s$} & \colhead{Density} & \colhead{Temperature} & \colhead{Pressure} & \colhead{Internal Energy} & \colhead{Helmholtz Free}  &\colhead{Entropy}\\
  ($a_0$) & (g$\,$cm$^{-3}$) & (K) & (GPa)~~~ & ~~~(Ha/el) & Energy (Ha/el)  & ~~($k_b$/el)
} 
\startdata
0.70 &  8.9658 & 5000 &  17713.9(3)  &$\;\,$   0.69251(3)   &$\;\,$   0.641237(7)  &  3.238(3) \\
0.80 &  6.0064 & 5000 &   8170.3(4)  &$\;\,$   0.41280(4)   &$\;\,$   0.351520(11) &  3.870(3) \\
0.90 &  4.2185 & 5000 &   4064.5(3)  &$\;\,$   0.24343(4)   &$\;\,$   0.173088(24) &  4.443(4) \\
1.00 &  3.0753 & 5000 &   2141.5(2)  &$\;\,$   0.13726(3)   &$\;\,$   0.058946(16) &  4.946(3) \\
1.10 &  2.3105 & 5000 &   1180.6(2)  &$\;\,$   0.06926(6)   &$-$0.016209(11) &  5.398(4) \\
1.20 &  1.7797 & 5000 &    675.0(1)  &$\;\,$   0.02513(5)   &$-$0.066864(21) &  5.810(4) \\
1.30 &  1.3998 & 5000 &    398.4(1)  &$-$0.00370(5)   &$-$0.101600(16) &  6.183(4) \\
1.40 &  1.1207 & 5000 &    242.1(1)  &$-$0.02265(8)   &$-$0.12587(3)   &  6.519(7) \\
1.50 &  0.9112 & 5000 &    151.8(2)  &$-$0.03527(6)   &$-$0.14310(4)   &  6.810(6) \\
1.60 &  0.7508 & 5000 &     98.0(1)  &$-$0.04400(6)   &$-$0.15565(2)   &  7.051(5) \\
1.86 &  0.4779 & 5000 &     38.2(1)  &$-$0.05755(7)   &$-$0.17565(3)   &  7.459(7) \\
2.00 &  0.3844 & 5000 &     25.74(8) &$-$0.06295(15)  &$-$0.18271(4)   &  7.563(12)\\
2.10 &  0.3321 & 5000 &     19.97(9) &$-$0.06627(13)  &$-$0.18655(4)   &  7.596(11)\\
2.20 &  0.2888 & 5000 &     15.51(9) &$-$0.0685(2)    &$-$0.19015(4)   &  7.683(16)\\
2.30 &  0.2528 & 5000 &     12.24(7) &$-$0.0702(2)    &$-$0.19312(5)   &  7.765(14)\\
2.40 &  0.2225 & 5000 &      9.64(6) &$-$0.0714(2)    &$-$0.19580(7)   &  7.855(17)\\
\enddata
\tablecomments{Table \ref{EOS} is published in its entirety in the 
electronic edition of the {\it Astrophysical Journal}.  A portion is 
shown here for guidance regarding its form and content.}
\end{deluxetable}

\begin{deluxetable}{cccl}
\tabletypesize{\scriptsize}
\tablecolumns{6}
\tablewidth{0pc}
\tablecaption{Coefficients of Free Energy Fit for the equation of state.\label{EOS_fit}}
\tablehead{
& $r_s$ & $T$ & ~~~~~coefficient\\
& (a.u.) & (a.u.) & ~~~~~~~(a.u.)\\
}
\startdata
$\frac{\partial F}{\partial T}$ & 1.75 & 0.001013 & $-$8.67942$\times 10^{-1}$\\
$F$                             & 1.75 & 0.001013 & $-$9.23942$\times 10^{-2}$\\
$F$                             & 1.75 & 0.003131 & $-$9.76575$\times 10^{-2}$\\
$F$                             & 1.75 & 0.006512 & $-$1.12223$\times 10^{-1}$\\
$F$                             & 1.75 & 0.011911 & $-$1.44120$\times 10^{-1}$\\
$F$                             & 1.75 & 0.020533 & $-$2.07646$\times 10^{-1}$\\
$F$                             & 1.75 & 0.034302 & $-$3.23222$\times 10^{-1}$\\
$F$                             & 1.75 & 0.056290 & $-$5.29287$\times 10^{-1}$\\
$F$                             & 1.75 & 0.091403 & $-$8.92177$\times 10^{-1}$\\
$F$                             & 1.75 & 0.147476 & $-$1.53056\\
$F$                             & 1.75 & 0.237021 & $-$2.64345\\
$F$                             & 1.75 & 0.380018 & $-$4.58672\\
$\frac{\partial F}{\partial T}$ & 1.75 & 0.380018 & $-$1.41575$\times 10^{+1}$
\enddata
\tablecomments{Table \ref{EOS_fit} is published in its entirety in the
  electronic edition of the {\it Astrophysical Journal}. A portion is
  shown here for guidance regarding its form and content. {\bf While
    this manuscript is still under review, a copy of the EOS fit as
    well as the computer interpolation program may be requested via
    email: militzer@berkeley.edu} }
\end{deluxetable}

\end{document}